\documentclass[review]{elsarticle}
\usepackage[margin=4cm]{geometry}

\usepackage{lineno,hyperref}

\usepackage{amsmath,amssymb,amsfonts}
\usepackage{algorithmic}
\usepackage{graphicx}
\usepackage{textcomp}
\usepackage{caption}
\usepackage{url}
\usepackage{booktabs} 
\usepackage{colortbl}
\usepackage[ruled, vlined, linesnumbered]{algorithm2e}

\usepackage{comment}
\usepackage{mathtools}
\usepackage{multirow}
\usepackage[dvipsnames]{xcolor}
\usepackage{balance}
\usepackage{bm}
\usepackage{subcaption}
\usepackage{eucal}
\def \red{\textcolor{black}}

\journal{Barin-Computer Interface-Special Issue}

\begin{document}

\begin{frontmatter}

\title{Brain2Object: Printing Your Mind from Brain Signals with Spatial Correlation Embedding}

\author{Xiang Zhang$^\P$, Lina Yao$^\P$, Chaoran Huang$^\P$\\ Salil S. Kanhere$^\P$, Dalin Zhang $^\P$, Yu Zhang$^\S$}
\address{$^\P$University of New South Wales, Sydney, Australia\\
$^\S$Stanford University, CA, USA}
\fntext[myfootnote]{xiang.zhang3@student.unsw.edu.au\\
\{lina.yao, chaoran.huang, salil.kanhere\}@unsw.edu.au\\yzhangsu@stanford.edu\\
Word Count: 9287
}




\begin{abstract}
Electroencephalography (EEG) signals are known to manifest differential patterns when individuals visually concentrate on different objects. In this work, we present an end-to-end digital fabrication system, {\it Brain2Object}, to print the 3D object that an individual is observing by decoding visually-evoked brain signals. We propose a unified training framework that combines multi-class Common Spatial Pattern and Convolutional Neural Networks to support the backend computation. We learn the dynamical graph representations of brain signals to accurately capture the structural information among EEG channels.
 A user-friendly interface is developed as the system front end. 
{\it Brain2Object} presents a streamlined end-to-end workflow that can serve as a template for deeper integration of BCI technologies to assist with our routine activities.

The proposed system is evaluated extensively using offline experiments and through an online demonstrator. The experimental results \red{show that our approach can achieve the recognition accuracy of 92.58\% on a benchmark dataset and 75.23\% on a locally collected dataset.} Moreover, our method consistently outperforms a wide range of baseline and state-of-the-art approaches. The proof-of-concept corroborates the practicality of our approach and illustrates the ease with which such a system could be deployed.
\end{abstract}


\end{frontmatter}

\section{Introduction}
An Electroencephalography (EEG)-based Brain-Computer Interface (BCI) enables people to communicate with the outside world by interpreting the EEG signals of their brains to interact with external devices such as wheelchairs and intelligent robots \cite{tomida2015active,zhang2017intent}. BCI systems have been widely studied for various real-world applications ranging from the healthcare domain \cite{pinheiro2016wheelchair,kaya20141d} to the entertainment industry \cite{Alomari2014}. 
\red{The availability of portable and affordable EEG collection devices,
 has opened up new opportunities for developing BCI applications on artworks such as musical composition \cite{hamadicharef2010brain,pinegger2017composing} and painting \cite{funk2013brain,george2010brain}. For instance, Pinegger {\it et al.} \cite{pinegger2017composing} develop a BCI system to compose music based on P300 patterns.}
Such {\it in-situ} use of this technology necessitates a shift towards a non-invasive way to collect EEG signals\cite{volker2018deep} (also employed by the aforementioned portable devices) as opposed to invasive approaches which rely on inserting electrodes into the scalp \cite{jayakar2016diagnostic}. 

However, one major challenge faced by the non-invasive collection of EEG signals is low signal-to-noise ratio (SNR) \cite{zhang2018converting}. This is caused by internal and external effects. The former include the loss in the strength of the signals as they pass through the thick medium of the skull, lack of concentration from the individual, etc. The latter includes the impact of environmental noise, light stimuli, and fidelity of acquisition device.
As a result, EEG signals inherently lack sufficient spatial resolution and insight on activities of deep brain structures. To this end, several studies have focused on denoising the signals and dealing with the aforementioned artifacts by statistical feature extraction ({\it e.g.}, principal component analysis, 
autoregressive model, wavelet analysis) \cite{shaw2016statistical,michelmann2018data,li2015autoregressive,zhang2017classification} and deep learning ({\it e.g.}, recurrent neural networks, autoencoder) \cite{li2017emotion,liu2017multi,li2015feature,lin2016classification}. 
\red{
The connectivity among brain electrodes reveals important information of brain states \cite{simpson2011exponential,agosta2013brain}. Different connections and link strengths may lead to different brain patterns, {\it i.e.}, brain networks. 
However, most of the existing studies have not fully taken the topology of the brain functional network into account.
}

In this work, we propose a unified approach, by learning the robust EEG representations through graphical brain networks, to recognize the imagery of an object seen by the individual. We first design a multi-class Common Spatial Pattern (CSP) for distilling the compact representations. CSP has proven success in extracting features using eigendecomposition based on the variance ratio between different classes \cite{elisha2017eeg}. 
Next, we propose Dynamical Graph Representation (DGR) of EEG signals \red{to adaptively embed the spatial relationship among the channels} (each channel represents one EEG electrode) and their neighbors by learning a dynamic adjacent matrix. Finally, a Convolutional Neural Network (CNN) is employed to aggregate higher-level spatial variations from the transformed graph representations.

On top of the abovementioned computational framework, we present a mind-controlled end-to-end system called {\it Brain2Object}. It enables an individual to print a physical replica on an object that s/he is observing by interpreting visually evoked EEG signals in a real-time manner. To enable the end-to-end workflow, the proposed system gathers the user's brain activities through EEG acquisition equipment and forwards the collected EEG data to a pre-trained model which automatically recognizes the object that the user is observing. 
\red{Imagine that a child observes a toy, for example Pinkie Pie (from {\it My Little Pony}), belonging to her friend. She likes it very much and wishes that she could have one too. Brain2Object can make her wish a reality by translating her brain signals, when she stars at the toy, to command the 3D printer to fabricate a copy\footnote{The related ethical issues such as IP and copyright of 3D models are not the focus of this work.}.}
 The ability to print a replica model of any observable object could be of tremendous value to a variety of professionals including engineers, artists, construction workers, etc.
To summarize, this paper makes the following key contributions: 
\begin{itemize}
    \item We present an end-to-end digital fabrication system, {\it Brain2Object}, atop of the precise decoding of human brain signals that allows an individual to instantly create a real-world replica (or model) of any object in her gaze. The proposed system is able to learn an illustration of an object seen by an individual from visually-evoked EEG signals, and print a model in real-time by automatically instructing a wireless connected 3D printer.
    \item We propose an effective EEG decoding model by learning a dynamical graph representation, which could adaptively embed structured EEG spatial correlations during the training process. A convolutional neural network is integrated for capturing discriminative feature representations as well as the intrinsic connections among the various EEG channels.
    \item The proposed approach is evaluated over a large scale benchmark dataset and a limited but locally collected dataset. Our method outperforms a wide range of baselines and state-of-the-art approaches in both instances, thus demonstrating the generalized nature of the approach. Finally, a prototype implementation demonstrates the practicality of {\it Brain2Object}.
\end{itemize}


\begin{figure*}[t]
\centering
\includegraphics[width=\textwidth]{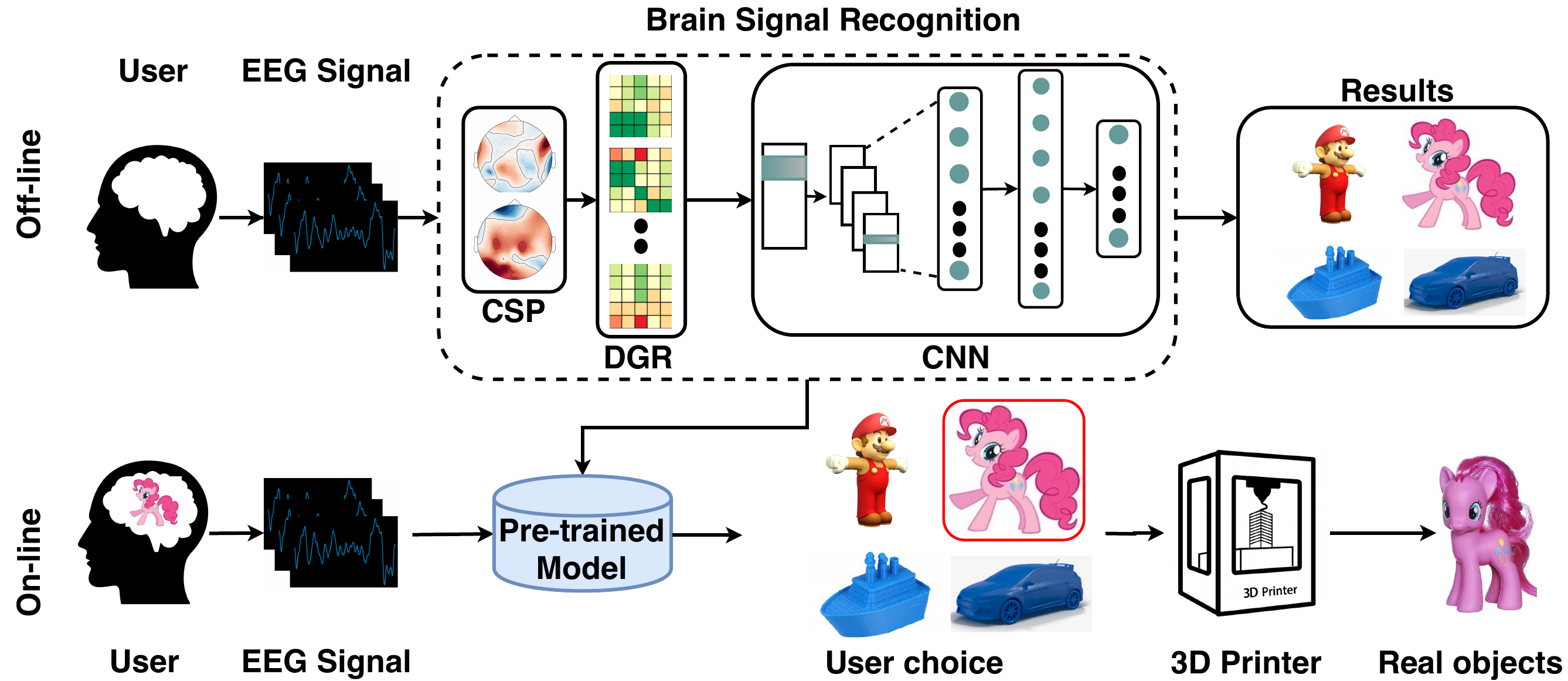}
\caption{Brain2Object overview. The object ({\it e.g.}, Pinkie Pie) observed by the user is reflected in the visually evoked EEG signals, which can be accurately recognized by the pre-trained recognition model.
The recognition module employs multi-class CSP for separating the multivariate signals into additive subcomponents which have maximum differences. The spatial dependencies among processed data is extracted by DGR and then forwarded to the CNN for recognition. The schematic of the identified object is loaded from the model library of the 3D printer to fabricate a replica..}
\label{fig:overview}
\vspace{-3mm}
\end{figure*}
\section{The Proposed System} 
\label{sec:the_proposed_system}

\subsection{System Overview} 
\label{sec:system_overview}
The overall aim of the proposed {\it Brain2Object} is to automatically recognize the object that the user desires to fabricate by analyzing one's visually-evoked brain signals and actuate a 3D printer accordingly. 

As shown in Figure~\ref{fig:overview}, the system is composed of an offline and online component.
The offline component aims to build a robust and effective unified classification model that can recognize the specific object that the user is observing by analyzing the corresponding brain signals that are evoked during this process.
We first record the EEG signals of individuals while they staring at an object (the details will be introduced in Section~\ref{sec:data_acquisition}). Next, the gathered EEG data are analyzed using multi-class CSP \cite{chin2009multi} to extract the eigenvalues of various categories of objects. CSP has been widely used in EEG signal analysis (such as motor-imagery brain activity classification \cite{kang2014bayesian,wang2006common}) and achieves comparable performance. Thus, in our system, we adopt CSP for discriminative spatial filtering to enhance the SNR of EEG signals. 
Through CSP, the EEG signals are mapping to a latent space where the inter-category variance difference is maximized.
Moreover, \red{to dynamically capture the structural information among EEG channels, we propose DGR to transform the CSP processed signals to a new space since graph representation has been shown to be helpful in refining and capturing spatial information \cite{song2018eeg}.}  
In the embedded space which encompasses the topological structure among the \red{brain regions}, each channel not only represents the amplitude of the measured signals but also reflects the dependency with other channels.
CNNs are widely used for processing of two-dimensional data in applications such as image recognition \cite{ciresan2011convolutional}, ubiquitous \cite{ning2018deepmag}, and object searching \cite{ren2017faster}, due to their salient features such as regularized structure, good spatial locality, and translation invariance.
Thus, we employ CNN as a classifier to distinguish the graph embedded features. After a number of training epochs, the converged classification model with stable weights is stored for online recognition.

During online operation, the user wears an EEG signal acquisition equipment. The equipment will collect her brain signals in real-time while she concentrating on a physical object. The gathered signals are forwarded to the pre-trained model to recognize the object. For example, as shown in Figure~\ref{fig:overview}, while the user focuses on `Pinkie Pie', the pre-trained model is empowered to recognize the `Pinkie Pie' and send an appropriate command to the 3D printer, which loads the 3D physical model and fabricates a copy.

\subsection{Multi-class Common Spatial Filtering} 
\label{sub:common_spatial_pattern}
CSP is widely used in the BCI field to find spatial filters that can maximize the variance between classes \cite{meisheri2018multiclass}. It has been successfully used for recognition of movement-related EEG \cite{ramoser2000optimal}. CSP was first introduced in binary classification problems but has since been extended to multi-class scenarios. In this paper, we adopt the one-vs-others strategy for multi-class CSP analysis. Please refer more details to \cite{meisheri2018multiclass,ang2012filter}.

Denote the gathered EEG data by $\mathbb{E} = \{\bm{E}_i, i \in 1, 2, \cdots, N\}$ where $N$ denotes the number of samples and each sample can be denoted by $\bm{E}_i \in \mathbb{R}^{M \times L}$ where $M$ is the number of EEG channels and $L$ denotes the number of time slices/points. Let $\bm{\bar{E}}_i$ be the output of CSP and then flow to downstream components for graphical feature learning.

\begin{figure}[t]
    \centering
    \begin{subfigure}[t]{0.4\textwidth}
        \centering
        \includegraphics[width=0.8\textwidth]{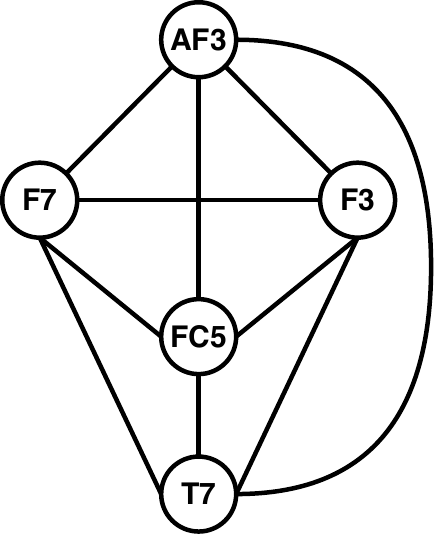}
        \caption{Graph}
    \end{subfigure}%
    \begin{subfigure}[t]{0.5\textwidth}
      \centering
      \includegraphics[width=0.8\textwidth]{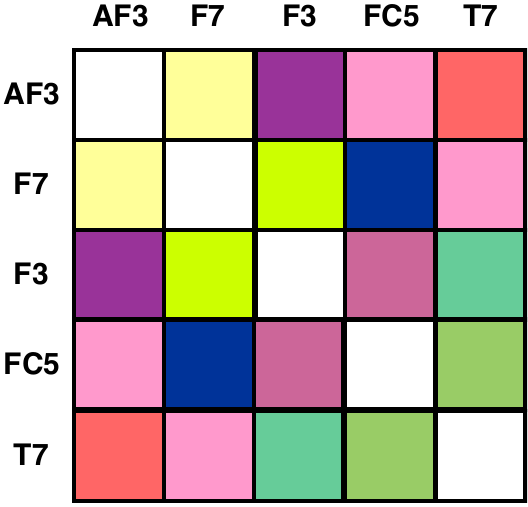}
      \caption{Adjacency matrix}
    \end{subfigure}%
    \caption{Illustration of a weighted undirected graph with 5 EEG electrodes and the corresponding adjacency matrix. The five vertices are reading from Frontal (F) and Temporal (T) lobes of human brain. The adjacency matrix is symmetric matrices, in which the colors denote the connection weights. \red{Please note this is just a schematic diagram to show how EEG electrodes connected in graph structure.}}
    \label{fig:graph}
\end{figure}

\subsection{Dynamical Graph Representation} 
\label{sub:dynamical_graph_representation}
In post-CSP processed EEG data $\bar{\bm{E}}$, each channel (row) separately provides the voltage amplitude of a specific electrode ({\it e.g.}, the value of channel $F3$ reflects the amplitude of EEG electrode $F3$) instead of the aggregated spatial information. The signals are discrete and discontinuous in the spatial domain. Hence, traditional spatial feature representation methods such as CNN are not well suited for further processing \cite{song2018eeg}. Instead, we invoke the knowledge of the connections of the brain neurons to map 
$\bar{\bm{E}}$ to a new space where each element represents not only the specific channel amplitude but also the spatial relationship with its neighboring channels. 

For this purpose, we regard the brain network as a complete weighted undirected graph with $M$ vertices where each vertex denotes a channel. The term `complete' denotes each vertex is connected to all the residual vertices in this graph. The graph can be defined as $\mathcal{G} = \{\mathcal{V}, \mathcal{E}, \mathcal{A}\}$ where $\mathcal{V} \in \mathbb{R}^M$ denotes the set of vertex with the number of $|\mathcal{V}| = M$ and $\mathcal{E}$ denotes the set of edges connecting the vertices. Suppose $\mathcal{A} \ \in \mathbb{R}^{M\times M}$ denotes the adjacency matrix representing the connectivity within $\mathcal{V}$. In particular, the element in the $i$-th row and $j$-th column of the adjacency matrix measures the weight or importance of the edges between the $i$-th and the $j$-th vertices.

During the training, the adjacency matrix is adaptively updated with the evolution of the model, which refers to the graph representation is dynamic. Hence, the name, Dynamic Graph Representation (DGR).
Figure~\ref{fig:graph} illustrates an example of a complete weighted undirected graph which is composited by five vertices which are reading from Frontal (F) and Temporal (T) lobes of the human brain. The diagonal elements are zero since each vertex is not connected to itself. 
However, the proposed representation should also contain information representative of each individual vertex. To incorporate this information, we add self-loop on the graph, which means integrate an identity matrix $\bm{I}$ to the adjacency matrix. The resulting DGR is thus represented as 
\begin{equation}\red{\bm{E}' = (\mathcal{A}+\bm{I})\sigma(\bm{w}\bar{\bm{E}} + \bm{b}) }\end{equation}
\red{where $\bm{w}$ and $\bm{b}$ denote transformation weights and biases and $\sigma$ denotes a non-linear activation function. 
}
The represented data $\bm{E}'$ with shape $[M, L]$ can dynamically learn the intrinsic relationship between different EEG channels by training a neural network and thus benefit most from discriminative EEG feature extraction.

\subsection{Convolutional Neural Networks} 
\label{sub:convolutional_neural_networks}
The DGR representation of the EEG signals serves as input to a specified CNN structure for feature refining and classification. CNN could capture the distinctive dependencies among the patterns associated with different EEG categories. The designed CNN comprises of one convolutional layer followed by three fully-connected layers (as shown in Figure~\ref{fig:overview}). The convolutional layer contains a set of filters to convolve the EEG data followed by the nonlinear transformation to extract the geographical features.
The input $\bm{E}'$ has shape $[M, L]$ with depth as $1$. We choose $D$ convolutional filters with size $[2,2]$ and stride size $[1, 1]$. The stride denotes the x-movements and y-movements distance of the filters. The same shape `zero padding' is used, which keeps the sample shape constant during the convolution calculation. In the convolutional operation, the feature maps from the input layer are convolved with the learnable filters
and fed to the activation function to generate the output feature map. For a specific convolutional area $\bm{x}$ which has the same shape as the filter, the convolutional operation can be described as
\begin{equation}\bm{x}' = tanh(\sum_{i}\sum_{j}\bm{f}_{ij}*\bm{x}_{ij})  \end{equation}
where $\bm{x}'$ denotes the filtered results while $\bm{f}_{ij}$ denotes the $i$-th row and the $j$-th column element in the trainable filter. We adopt the widely used tanh activation function for nonlinearity.
The depth of EEG sample transfers to $D$ through the convolutional layer and the sample shape is changed to $[M, L, D]$. The features learned from the filters are concatenated and flattened to $[1, M*L*D]$ and forwarded to the first fully-connected layer. Thus, the first fully connected layer has $M*L*D$ neurons, after which, the second and the third (the output layer) fully-connected layers have $D'$ and $K$ neurons, respectively. The operation between the fully-connected layers can be represented by 
\begin{equation}\bm{E^{h+1}} = softmax(\bar{\bm{w}}\bm{E^{h}} + \bar{\bm{b}})  \end{equation}
where $h$ denotes the $h$-th layer and $\bar{\bm{w}}$, $\bar{\bm{b}}$ denote the corresponding weights matrix and biases.
The softmax function is used for activation. For each EEG sample, the corresponding label information is presented by one-hot label $\bm{y} \in \mathbb{R}^K$. The error between the predicted results and the ground truth is evaluated by cross-entropy
\begin{equation}loss = - \sum_{k=1}^{K}\bm{y}_klog(p_k)  \end{equation}
where $p_k$ denotes the predicted probability of observation of an object belonging to category $k$.
The calculated error is optimized by the AdamOptimizer algorithm. To minimize the possibility of overfitting, we the dropout strategy and set the drop rate to 50\%.

\red{At last, it is worth noting that deep learning networks has been demonstrated able to remove artifacts from the contaminated EEG signals due to the excellent latent feature learning ability \cite{yang2018automatic}.
Thus, in this work, the deep learning architecture including DGR and CNN are also be used to reduce the impact brought by artifacts.}

\begin{figure}[t]
\centering
\includegraphics[width=0.7\textwidth]{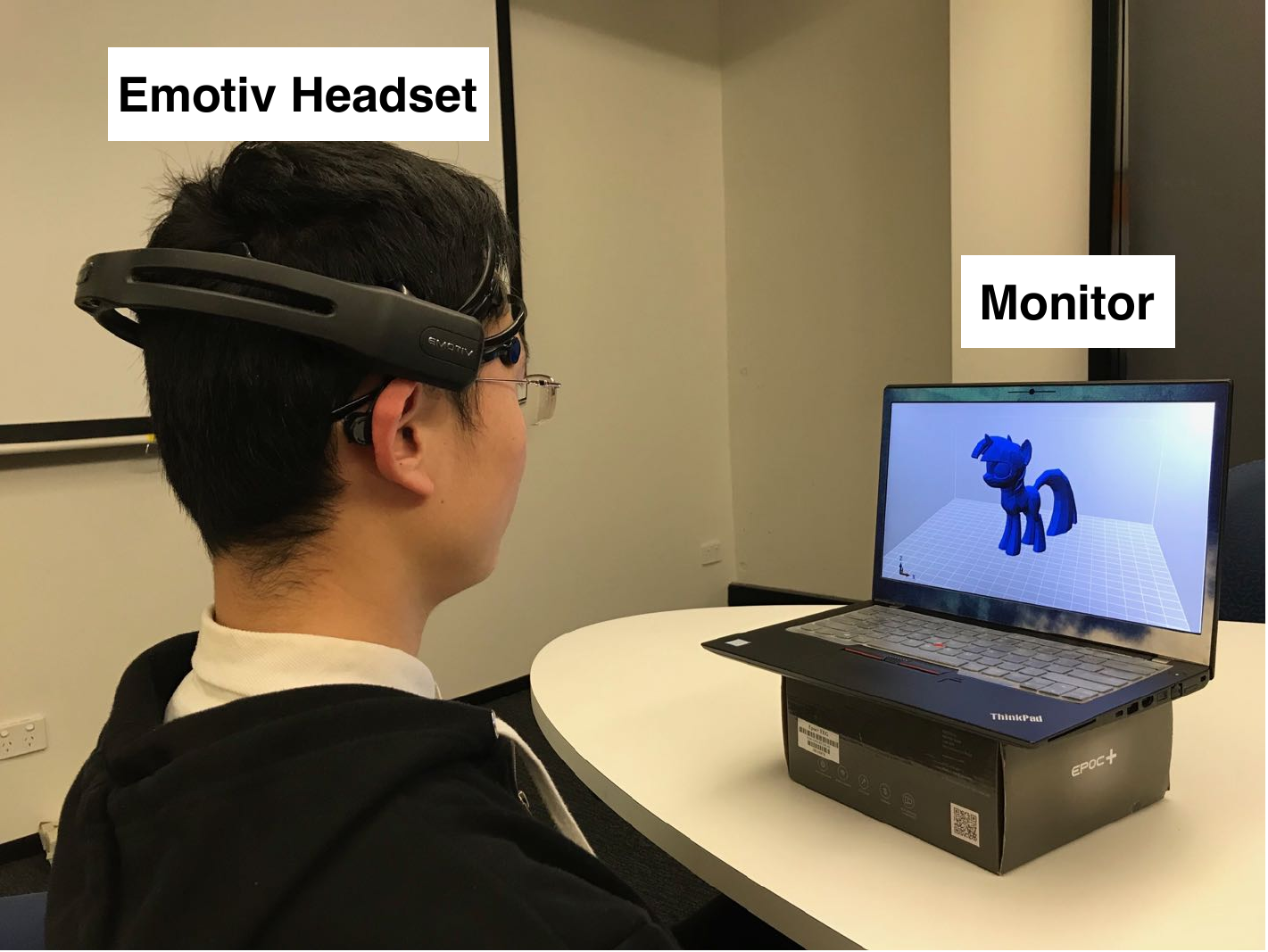}
\caption{Data acquisition experiment. The participant wears the EPOC+ Emotiv headset with 14 channels siting in front of a monitor which shows the Pinkie Pie.
 }
\label{fig:experiment}
\vspace{-3mm}
\end{figure}
\section{Data Acquisition} 
\label{sec:data_acquisition}
In this section, we gather a local EEG dataset which reflects the user's brain voltage fluctuation under visual stimulation of a number of object images.
\red{As a preliminary study,} we limit our study to include images of 4 objects: a car, a boat, the Pinkie Pie Pony, and the Mario (from the video game).

We recruit 8 healthy participants (aged 22-27 years) including 5 males and 3 females to participate in this study. \red{The experiment is approved by the UNSW Human Research Ethics Committee\footnote{https://research.unsw.edu.au/recs/human-research-ethics-home} and the project number is HC190315.} The data collection is conducted in a quiet room. As shown in Figure~\ref{fig:experiment}, the subject wears the EPOC+ Emotiv EEG headset which contains 14 channels (\red{AF3, F7, F3, FC5, T7, P7, O1, O2, P8, T8, FC6, F4, F8, and AF4}) corresponding to the 10-20 system (which is an internationally recognized method to describe and apply the location of scalp electrodes). The sampling rate is set as 128 Hz
and the headset can wireless connection with the computer over Bluetooth. The participants sit in a comfortable armchair, maintain a relaxed composure and gaze at a monitor placed approximately 0.5 meters in front of them.
Each subject participates in 10 sessions and each session contains 4 trials.

Each trial lasts for 15 seconds and is comprised of three phases, each lasting 5 seconds. In the first phase, the monitor shows an empty slide and the subject is asked to be relaxed. In the second phase, 
a random object picture is presented in the middle of the screen and the subject is instructed to focus on the projected image. The final phase is identical to the first phase. \red{The third phase is designed to leave subject more rest time to relieve fatigue caused by concentration.} Naturally, only EEF signals collected during the second phase are used in our dataset. In the second phase, the image is chosen with equal probability from the 4 aforementioned images.
To keep the balance of the dataset, the final EEG data of each specific participant is composed of 40 trials where each object appears 10 times. 

As a result, for each subject, there are 40 trials where each trial lasts for 5 seconds. Hence, each participant contributes 200 seconds of EEG signals. Since the sampling rate is 128 Hz, each subject contributes $25,600 = 128\times 200$ sampling points, which means the dataset has $204,800 $ sampling points in total.

\section{Experiments} 
\label{sec:experiments}

\subsection{Datasets} 
\label{sub:datasets}
In Section~\ref{sec:data_acquisition}, we introduced data acquisition experiments that were conducted locally in our lab. However, this local dataset has several drawbacks: 1) the spatial and temporal resolutions are limited since the headset used only has 14 channels and a low sampling frequency of 128 Hz.; 2) the fidelity (measured by SNR) of the signals is rather low because the resolution of recording the voltage is $0.51V$, which is significantly lower than the sophisticated devices such as those used for medical studies ({\it e.g.}, BCI2000); 3) the limited (8) number of participants in our dataset. A dataset containing a wider population of subjects is necessary for effective evaluations. \red{To demonstrate the reliability of our method, we also adopt a widely used public dataset that has more subjects and higher resolution in the experiments. In the rest of this paper, the local dataset is named EEG-L while the public dataset is named EEG-P\footnote{\red{The EEG-L is used to demonstrate that our method can decode EEG signals evoked by visual stimulus while EEG-P is only used to show the reliability of our method. Moreover, the 3D printing system proposed in this paper is triggered by visual stimulation rather than motor imagery.}}.}
 The EEG-P (eegmmidb) is collected by the BCI200 EEG system which records the brain signals using 64 channels with a sampling rate of 160Hz\footnote{Please refer more details about this dataset, like trails/scenarios/subject, to the following link \url{https://www.physionet.org/pn4/eegmmidb/}}. EEG data is recorded while the subjects are provided a visual stimulus (on a monitor) of certain actions and asked to imagine performing those actions. The four actions (left hand, right hand, both hands, and both feet) are labelled from 1 to 4. \red{In order to reduce computing resources, we form a subset by randomly select 20 subjects from eegmmidb.} In our dataset, the 560,000 samples belonging to 4 different labels and 20 subjects are selected with each subject having 28,000 samples.  In EEG-L, the four objects (Mario, car, boat, and Pinkie Pie pony) are labelled as 1, 2, 3, 4, correspondingly. \red{During signal processing, we separately select 64 and 14 time slices for EEG-P and EEG-L to train CSP. The adjacency matrix in DGR is first initialized by Gaussian distribution $\mathcal{N}(0, 0.01)$ and then be dynamically updated through back-propagation. }

Both datasets are further \red{randomly} sub-divided into a training set and testing set \red{to avoid trail-wise biases}. The former comprises 80\% of the data, while the latter contains the remaining 20\%. The training set is split into 4 equal mini-batches ({\it i.e.}, each mini-batch has 20\% of the data). All the features are normalized by the z-score method. The normalization parameters are noted for use during the online phase. The segmentation time window is set to 64 and 16 for EEG-P and EEG-L, respectively. The rate of overlap is 50\% for both datasets. 


\begin{table}[]
\centering
 \caption{Overall comparison with state-of-the-art models and
}
\label{tab:overall_comparison}
\resizebox{\textwidth}{!}{
\begin{tabular}{llllll}
\hline
\textbf{Dataset} & \textbf{Method} & \textbf{Accuracy} & \textbf{Precision} & \textbf{Recall} & \textbf{F-1} \\ \hline
\multirow{11}{*}{\textbf{EEG-P}} & \textbf{KNN} & 0.6962 & 0.7325 & 0.7552 & 0.7437 \\
 & \textbf{RF} & 0.7137 & 0.7536 & 0.7328 & 0.7431 \\
 & \textbf{SVM} & 0.6692 & 0.7122 & 0.7156 & 0.7139 \\
 & \textbf{CSP+KNN} & 0.9134 & 0.9273 & 0.9135 & 0.9203 \\
 & \textbf{CNN} & 0.8638 & 0.8619 & 0.8722 & 0.8670 \\
& \textbf{Sturm \cite{sturm2016interpretable}} & 0.8327    &0.8556&    0.8559&    0.8557 \\
& \textbf{Yang \cite{yang2015use}} & 0.8631    &0.8725    &0.8669    &0.8697 \\
& \textbf{Park \cite{park2014augmented}} & 0.8915    &0.9013    &0.9125    &0.9069\\
& \textbf{Thomas \cite{thomas2017eeg}} & 0.7986    &0.8031&    0.8219&    0.8124\\  
&\textbf{Zhang \cite{zhang2018converting}} &0.8325&0.8261&    0.8433    &0.8346\\
 & \textbf{Ours} & \textbf{0.9258} & \textbf{0.9325} & \textbf{0.925} & \textbf{0.9287} \\ \hline
\multirow{11}{*}{\textbf{EEG-L}} & \textbf{KNN} & 0.5108 & 0.5212 & 0.5436 & 0.5322 \\
 & \textbf{RF} & 0.5826 & 0.6258 & 0.6246 & 0.6252 \\
 & \textbf{SVM} & 0.6538 & 0.6684 & 0.6825 & 0.6754 \\
 & \textbf{CSP+KNN} & 0.5833 & 0.5773 & 0.5833 & 0.5803 \\
 & \textbf{CNN} & 0.6863 & 0.7021 & 0.6038 & 0.6493 \\
 & \textbf{Sturm \cite{sturm2016interpretable}} & 0.6988    &0.7021    &0.7086&    0.7053 \\
& \textbf{Yang \cite{yang2015use}} & 0.5832&    0.5968    &0.6013    &0.5990 \\
& \textbf{Park \cite{park2014augmented}} & 0.6892    &0.6995    &0.7021&    0.7008 \\
& \textbf{Thomas \cite{thomas2017eeg}} & 0.6679&    0.6759    &0.6821    &0.6790 \\  
&\textbf{Zhang \cite{zhang2018converting}} &0.6731    &0.6889    &0.6921    &0.6905 \\
 & \textbf{Ours} & \textbf{0.7523} & \textbf{0.7602} & \textbf{0.7528} & \textbf{0.7564} \\ \hline
\end{tabular}
}
\end{table}

\begin{figure*}[!t]
    \centering
    \begin{subfigure}[t]{0.48\textwidth}
        \centering
        \includegraphics[width=\textwidth]{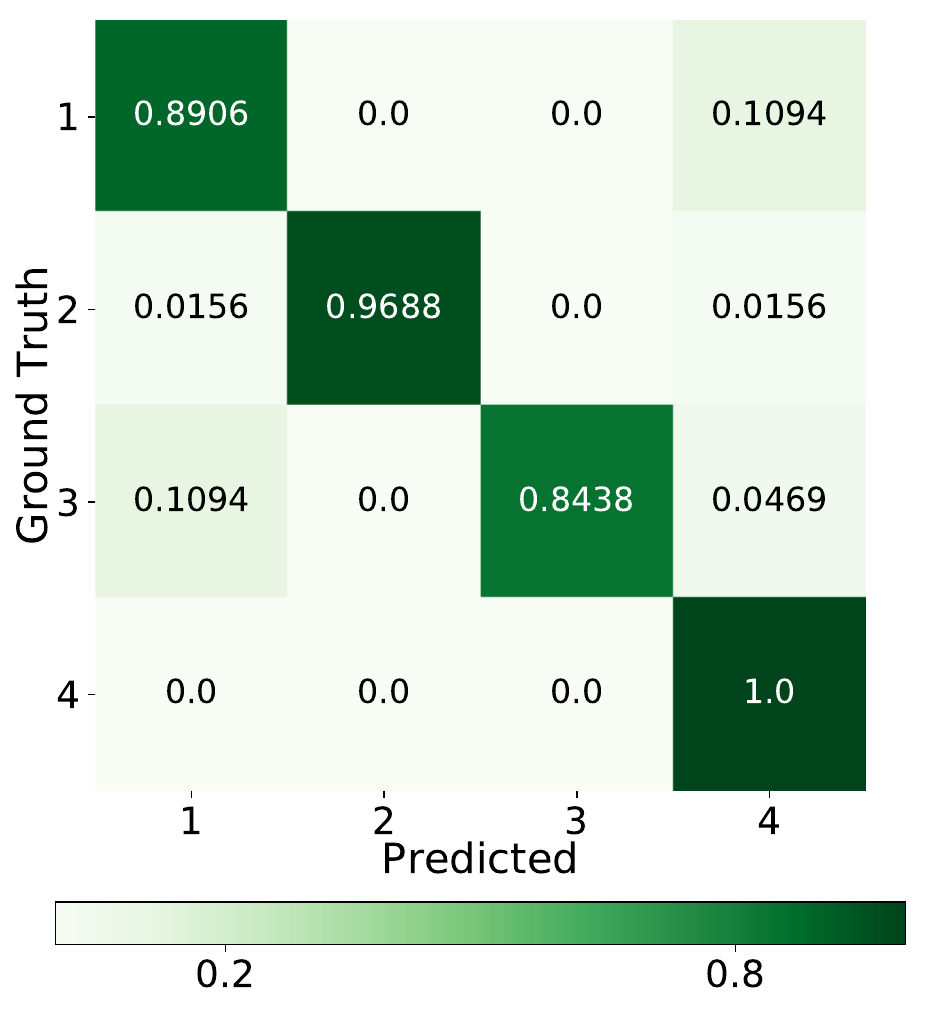}
        \caption{EEG-P confusion matrix.}
    \end{subfigure}%
    \begin{subfigure}[t]{0.48\textwidth}
      \centering
      \includegraphics[width=\textwidth]{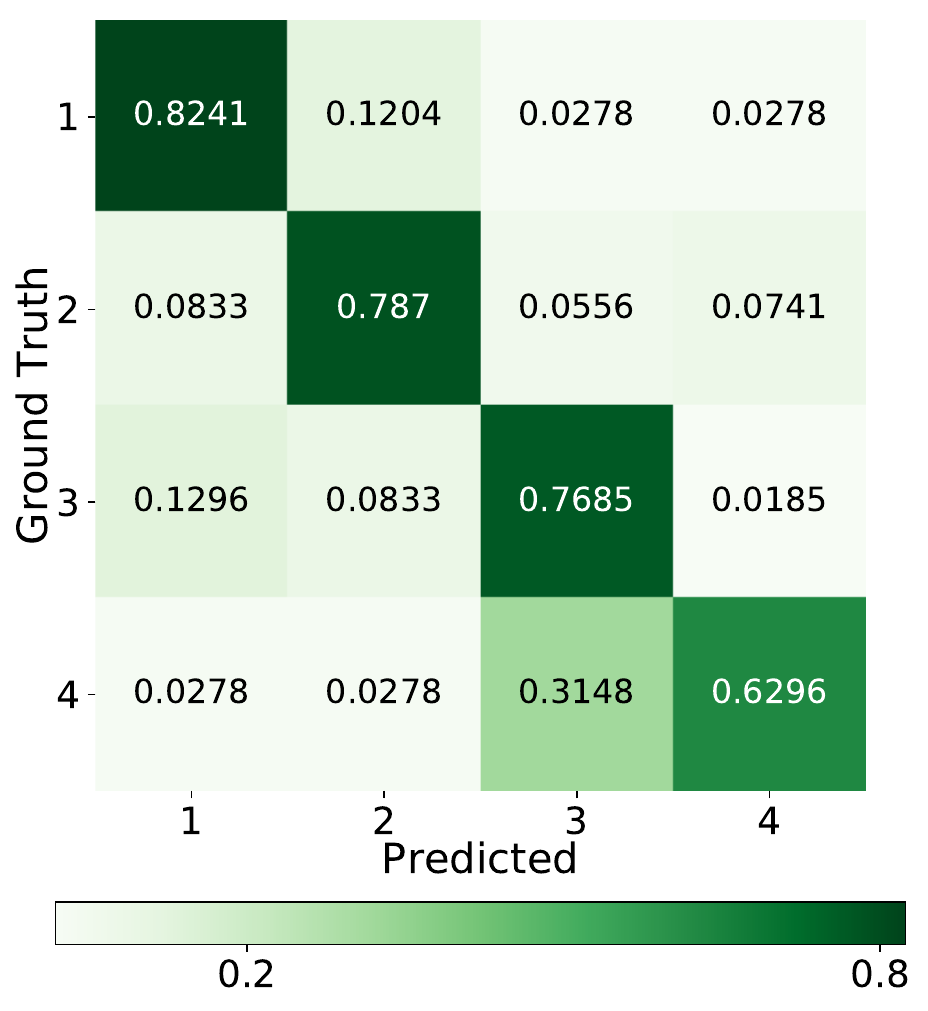}
      \caption{EEG-L confusion matrix.}
    \end{subfigure}%

    \begin{subfigure}[t]{0.5\textwidth}
        \centering
        \includegraphics[width=\textwidth]{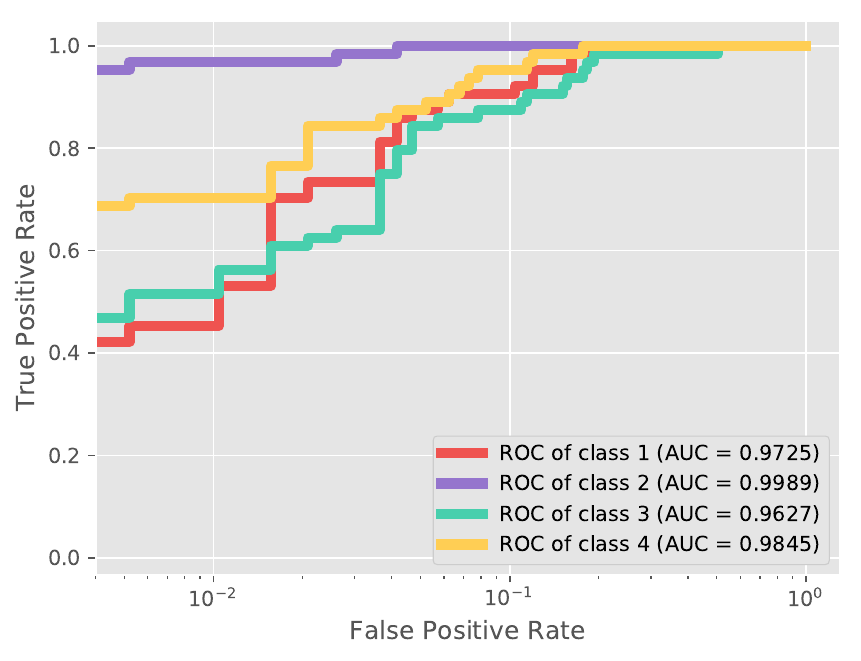}
        \caption{EEG-P ROC and AUC.}
    \end{subfigure}%
    \begin{subfigure}[t]{0.5\textwidth}
        \centering
        \includegraphics[width=\textwidth]{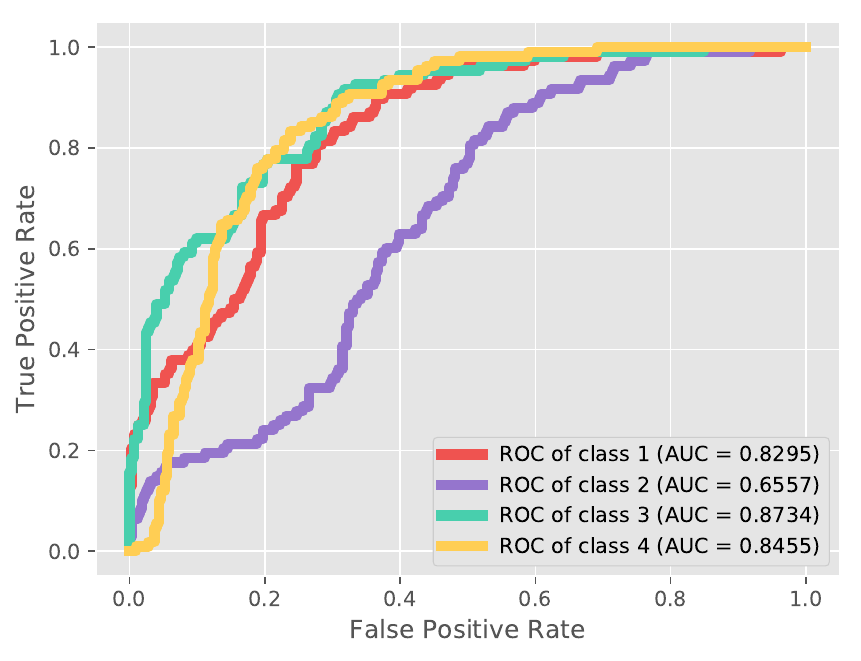}
        \caption{EEG-L ROC and AUC.}
    \end{subfigure}%
    \caption{Confusion matrix and ROC curves with AUC score.}
    \label{fig:cm_roc}
    \vspace{-3mm}
\end{figure*}

\subsection{Overall Comparison} 
\label{sub:overall_comparison}
Next, we report the performance of Brain2Object. Recall the adopted classification method combines the multi-class CSP, dynamic graph representation and the convolutional neural networks. We report the average performance over 5 runs.  All the experiments are run on the Titan X (Pascal) GPU. 

First, we provide the overall comparison with several widely used baselines including KNN, Random Forest (RF), Support Vector Machine (SVM). 
The key parameters of the baselines are listed here: KNN with 3 nearest neighbors; SVM with RBF kernel; RF with 50 trees. The independent CNN has the identical structure of the CNN component in our system as introduced in Section~\ref{sub:convolutional_neural_networks}. The kernel and stride information have been provided above, the learning rate is set as 0.0005 and the depth of convolutional layer $D$ equals to 10. The number of hidden neurons in the second fully-connected layer is 1000 for EEG-P and 120 for EEG-L. All the parameters are determined by empirical tuning. We also compare with a range of competitive state-of-the-art models:
\begin{itemize}
    \item Sturm {\it et al.} \cite{sturm2016interpretable} \red{propose the application of Deep Neural Networks with layer-wise relevance propagation for high-dimensional EEG analysis and classification. }
    \item Yang {\it et al.} \cite{yang2015use} \red{investigate the combination of augmented CSP and CNN for the aim of precise recognition of multi-class motor imagery brain signals.}
    \item Park {\it et al.} \cite{park2014augmented} \red{introduce an augmented complex-valued CSP based on the correlation between EEG channels in order to cater for general complex signals with noncircular probability distributions.}
    \item Thomas {\it et al.} \cite{thomas2017eeg} \red{aim to improve EEG classification performance} by carefully selecting the subject-specific spatial and spectral features. 
    \item Zhang {\it et al.} \cite{zhang2018converting} combine Recurrent Neural Networks (RNNs) with CNN to extract the temporal-spatial features from brain signals \red{for the aim of accurate classification of multi-class EEG signals.}
\end{itemize}
The results are depicted in Table~\ref{tab:overall_comparison}. \red{All the experimental results are measured on the same dataset. We report the average accuracy across all subjects.} It can be observed that our method achieves the highest accuracy (which corresponds to 0.9258 for EEG-P and 0.7523 for EEG-L) in comparison with state-of-the-art approaches for both datasets. The experimental results indicate that the proposed method are effective in learning the latent high-level features among EEG signals.
One can also readily observe that all methods achieve lower accuracy for EEG-L and as compared to EEG-P. Some of the drawbacks of EEG-L were already highlighted in Section~\ref{sub:datasets} including low fidelity, poor spatial-temporal coverage, and equipment limitations. Another reason could be that our participants did not have extensive experience with the usage of EEG headsets and neither were there any specialized technicians available to assist. Finally, the emotional state of the participants may have also influenced the EEG signals.

Furthermore, we present a range of additional metrics for our approach. This includes the confusion matrix and ROC curves with AUC scores in Figure~\ref{fig:cm_roc} and precision, recall and F-1 score for each category in Table~\ref{tab:report}. 

\begin{table}[]
\centering
\caption{Classification report including precision, recall, and F-1 score
}
\label{tab:report}
\resizebox{0.6\textwidth}{!}{
\begin{tabular}{l|l|llll}
\hline
\multirow{2}{*}{\textbf{Dataset}} & \multirow{2}{*}{\textbf{Metrics}} & \multicolumn{4}{c}{\textbf{Category}} \\ \cline{3-6}
 &  & \textbf{1} & \textbf{2} & \textbf{3} & \textbf{4} \\ \hline
\multirow{3}{*}{\textbf{EEG-P}} & \textbf{Precision} & 0.88 & 1 & 1 & 0.85 \\
 & \textbf{Recall} & 0.89 & 0.97 & 0.84 & 1 \\
 & \textbf{F-1} & 0.88 & 0.98 & 0.92 & 0.92 \\ \hline
\multirow{3}{*}{\textbf{EEG-L}} & \textbf{Precision} & 0.77 & 0.77 & 0.66 & 0.84 \\
 & \textbf{Recall} & 0.82 & 0.79 & 0.77 & 0.63 \\
 & \textbf{F-1} & 0.8 & 0.78 & 0.71 & 0.72 \\ \hline
\end{tabular}
}
\end{table}
\begin{figure}[t]
\centering
\includegraphics[width=0.8\textwidth]{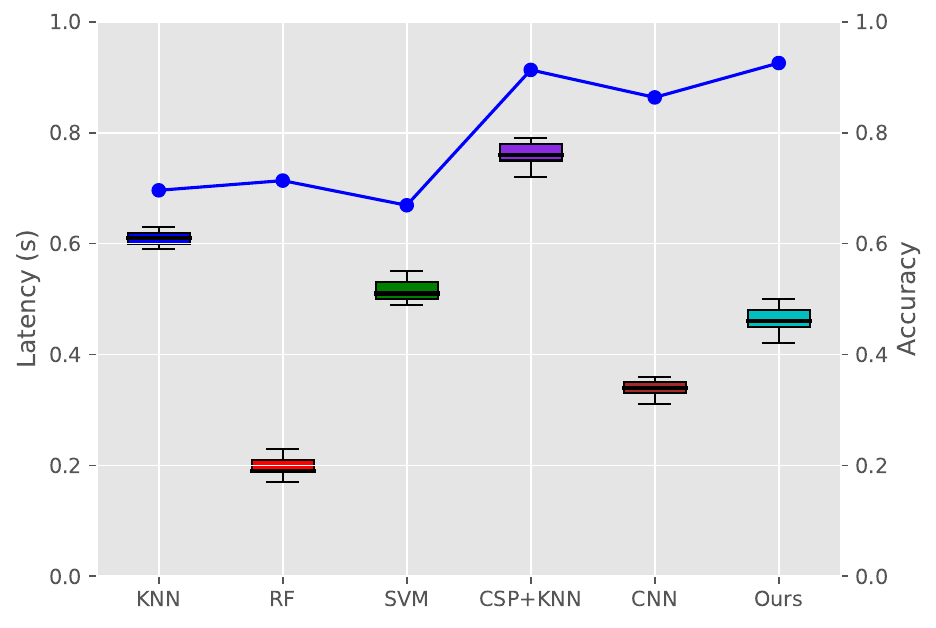}
\caption{Latency comparison against the accuracy. 
It can be observed that our approach achieves the highest accuracy with an acceptable latency.
}
\label{fig:latency}
\vspace{-3mm}
\end{figure}

\begin{figure}[t]
    \centering
    \begin{subfigure}[t]{0.45\textwidth}
        \centering
        \includegraphics[width=\textwidth]{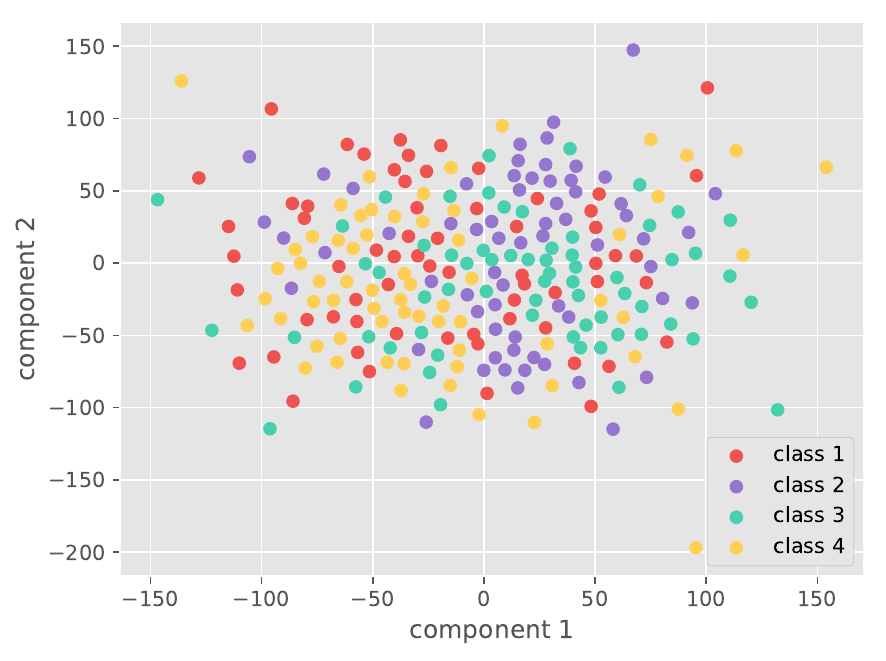}
        \caption{EEG-P raw data}
    \end{subfigure}%
    \begin{subfigure}[t]{0.45\textwidth}
      \centering
      \includegraphics[width=\textwidth]{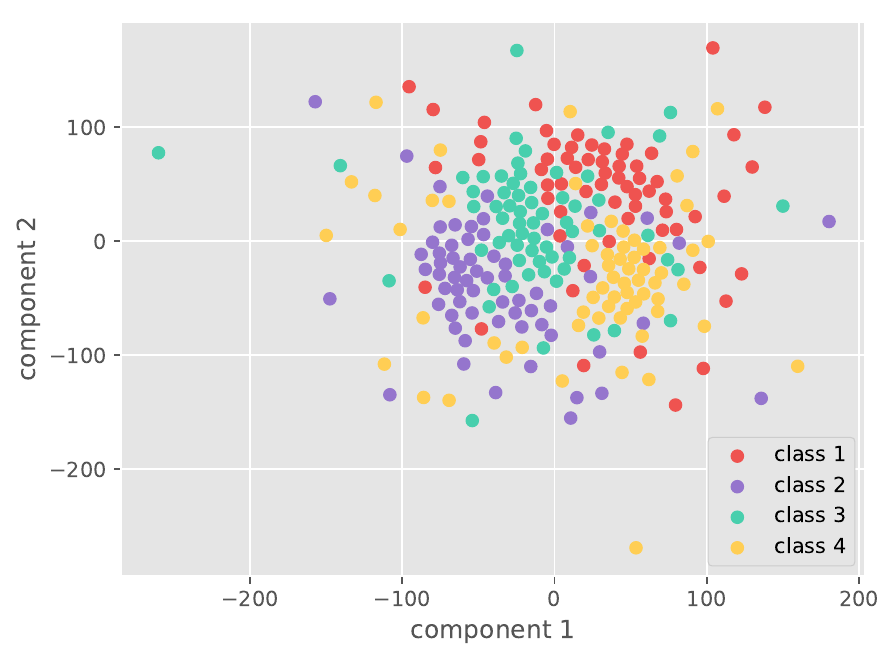}
      \caption{EEG-P feature}
    \end{subfigure}%

    \begin{subfigure}[t]{0.45\textwidth}
        \centering
        \includegraphics[width=\textwidth]{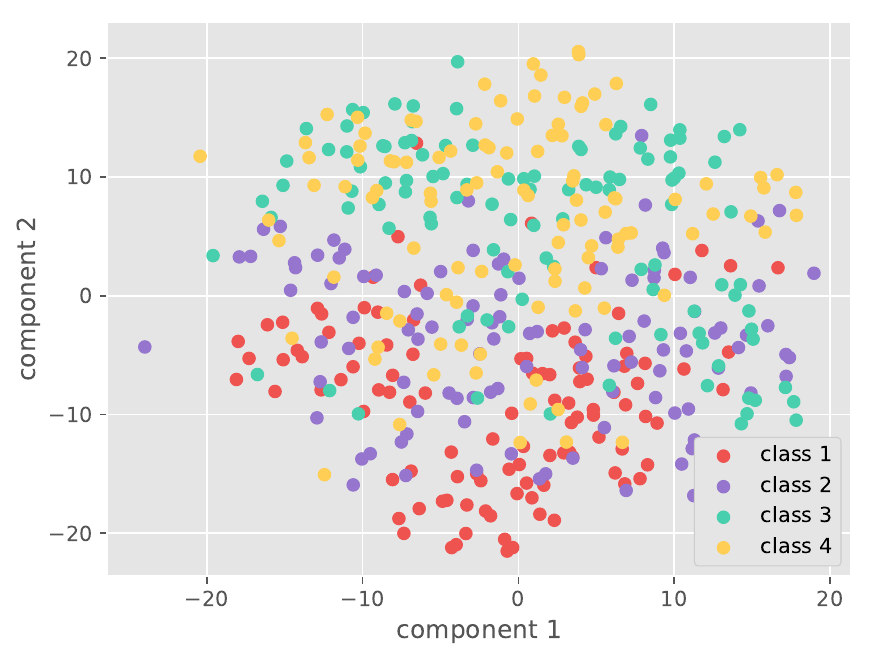}
        \caption{EEG-L raw data}
    \end{subfigure}%
    \begin{subfigure}[t]{0.45\textwidth}
        \centering
        \includegraphics[width=\textwidth]{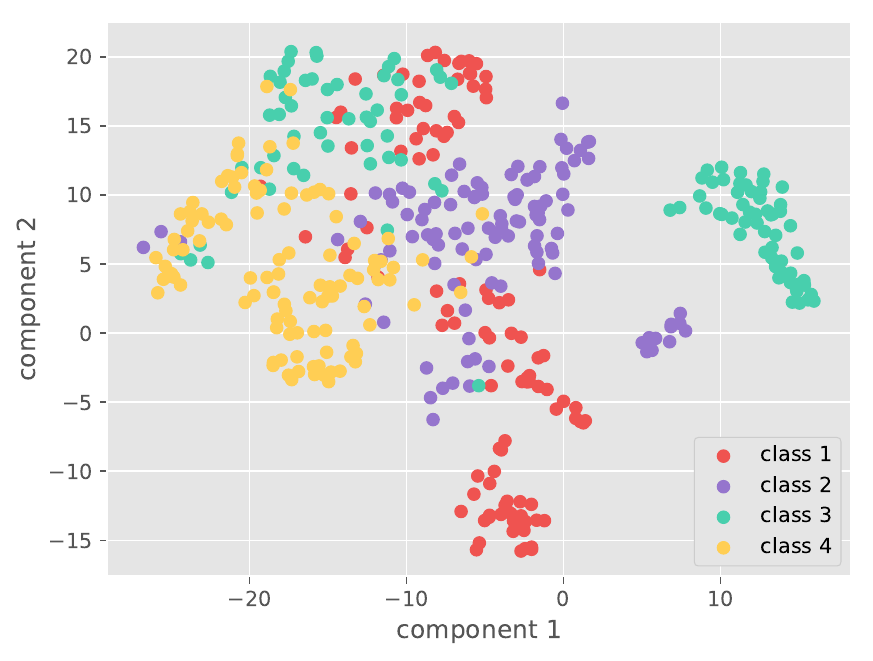}
        \caption{EEG-L feature}
    \end{subfigure}%
    \caption{A visualization comparing the raw data and extracted features for the two datasets. Both the raw data and the extracted feature are visualized from the corresponding testing set. This comparison demonstrates that our approach can (i) maximize the distance between the EEG data points and (ii) accurately extract the distinctive representations from the raw data.}
    \label{fig:visualization}
    \vspace{-3mm}
\end{figure}

\subsection{Latency} 
\label{sub:latency}
In addition to accuracy, latency is also an important performance metric for a system such as Brain2Object. \red{Here, latency refers to the time interval between our system receives the EEG signals and produces the classification results.}

Figure~\ref{fig:latency} illustrates the latency achieved by our method in comparison with a selected sub-set of baselines used in Section~\ref{sub:overall_comparison}.
We can observe that our approach has competitive latency compared with other methods while achieving the highest accuracy. The overall latency is less than 0.5 seconds. We computed the latency incurred by the different methods that are employed in our system and observed that CNN requires about 0.35 seconds for execution, while CSP and DGR together only require about 0.12 seconds. This illustrates that the use of deep learning techniques does not have a significant effect on the overall latency.

In a fully functional system, the end-to-end latency is not only comprised of the algorithmic latency but also includes the delay incurred for signal acquisition and signal transmission. The latter will be discussed in Section~\ref{sec:online_demonstration}. In the following, we evaluate signal acquisition latency. In the proposed system, the signal collection time is related to the acquisition equipment, in essential, the sampling rate. For BCI2000, a single sample/segment is composed of 64 time points, which is gathered in $0.4 = 64/160$ seconds with 160 Hz sampling frequency. On the other hand, the Epoc+ Emotiv headset only requires $ 0.11= 14/128$ seconds for signal collection. We can observe that the precise equipment can achieve higher accuracy but demand larger latency. In contrast, the off-the-shelf low fidelity headset has lower accuracy but also low latency. But this statement, which is similar to `no free lunch' rule, is based on the fact that our segment length equals the channel number. Could EEG-P keep the high-level accuracy with the decrease of channel amount in order to implement competitive performance with low latency at the same time? This meaningful scope deserves more attention in the future.

\subsection{Visualization} 
\label{sub:visualization}
To offer a different perspective into the performance of our system, we present a visualization of the data at two levels. At the system level, as a unified classification model, we visualize the raw EEG data and the extracted distinguishable features for comparison. \red{The features are taken from the second fully-connected layer in the CNN classifier because this layer has a direct relationship with the output layer and thus able to represent the quality of extracted features.} In Figure~\ref{fig:visualization}, the visualization of both EEG-P and EEG-L is presented. In which, Principal Component Analysis (PCA) is used for dimensionality reduction before visualization. Through the comparison, we can demonstrate that our approach maximizes the distance among EEG signals and has the ability to automatically extract the distinctive representations from raw data. 

\begin{figure}[t]
\centering
\includegraphics[width=0.8\textwidth]{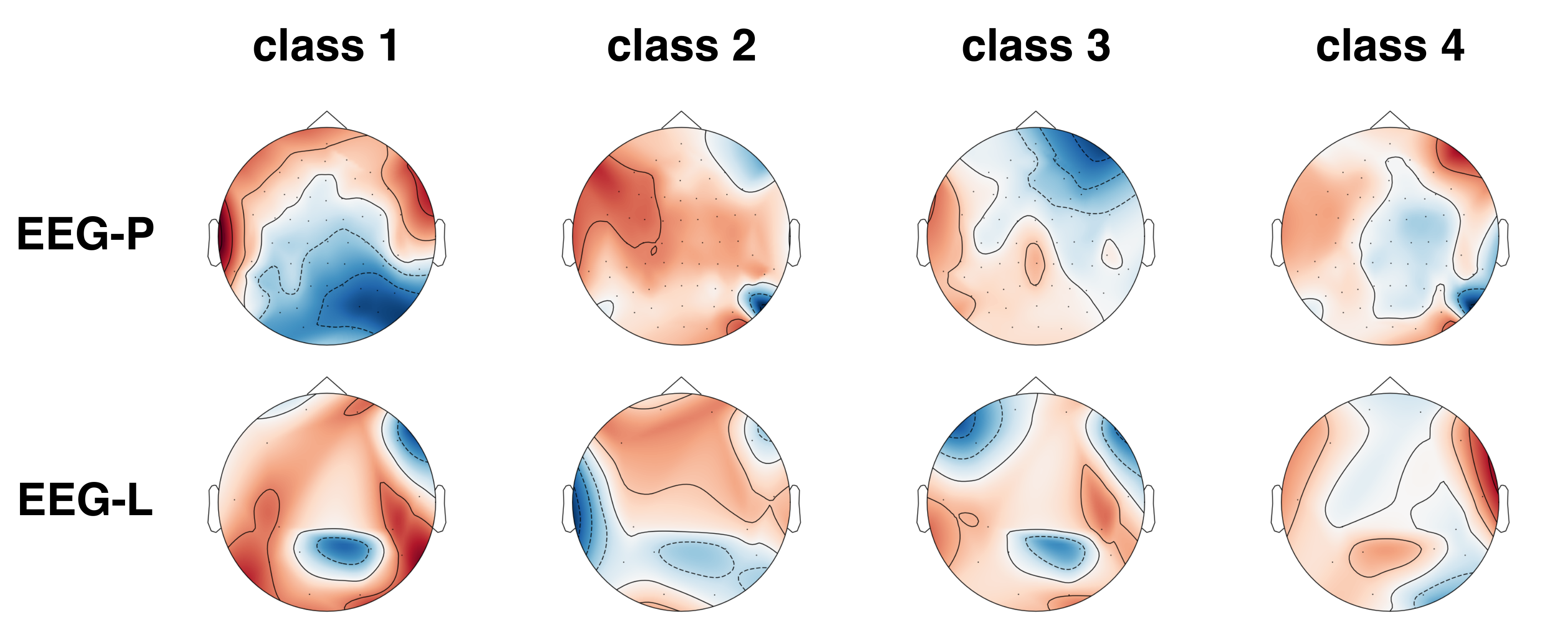}
\caption{Topography after CSP processing. Each topography in the first row contains 64 channels while the second row map contains 14 channels.
Through the comparison, it can be observed that the patterns belong to different categories are obviously variant, which indicates that the CSP processed features ought to be easier classified.
 }
\label{fig:topo}
\end{figure}

At the component level, we present the topography of various categories in each dataset. Figure~\ref{fig:topo} provides the EEG topographies after CSP processing. The first row represents the EEG-P dataset with 64 channels while the second row represents the EEG-L dataset with 14 channels. 
The channel names and positions strictly obey the international 10-20 system. Through the comparison, it can be observed that the patterns belong to different categories are obviously varying. This suggests that the CSP processed features ought to be classified easily.

\begin{figure}[t]
\centering
\includegraphics[width=0.7\textwidth]{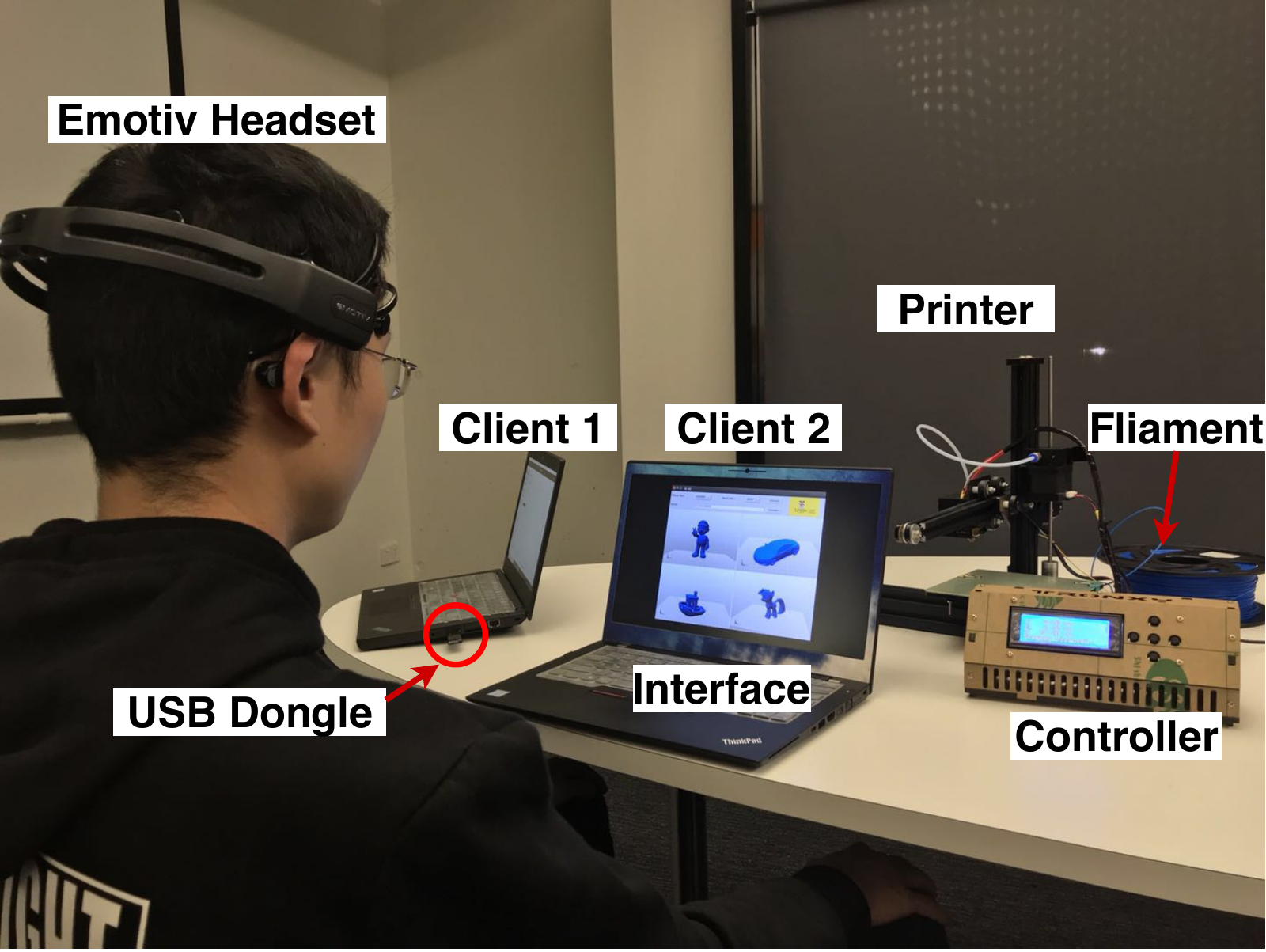}
\caption{Online testing scenario. The user's EEG signals are collected by Emotiv headset for recognition. The correspond object will be printed through the 3D printer.}
\label{fig:online_scenario}
\vspace{-3mm}
\end{figure}

\begin{figure}[t]
\centering
\includegraphics[width=0.7\textwidth]{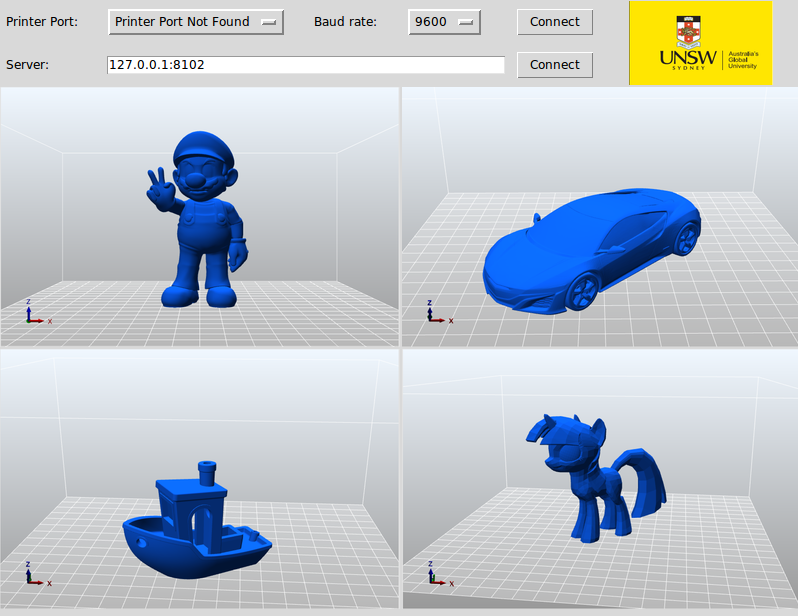}
\caption{User Interface.
}
\label{fig:GUI}
\end{figure}

\begin{figure}[t]
\centering
\includegraphics[width=0.8\textwidth]{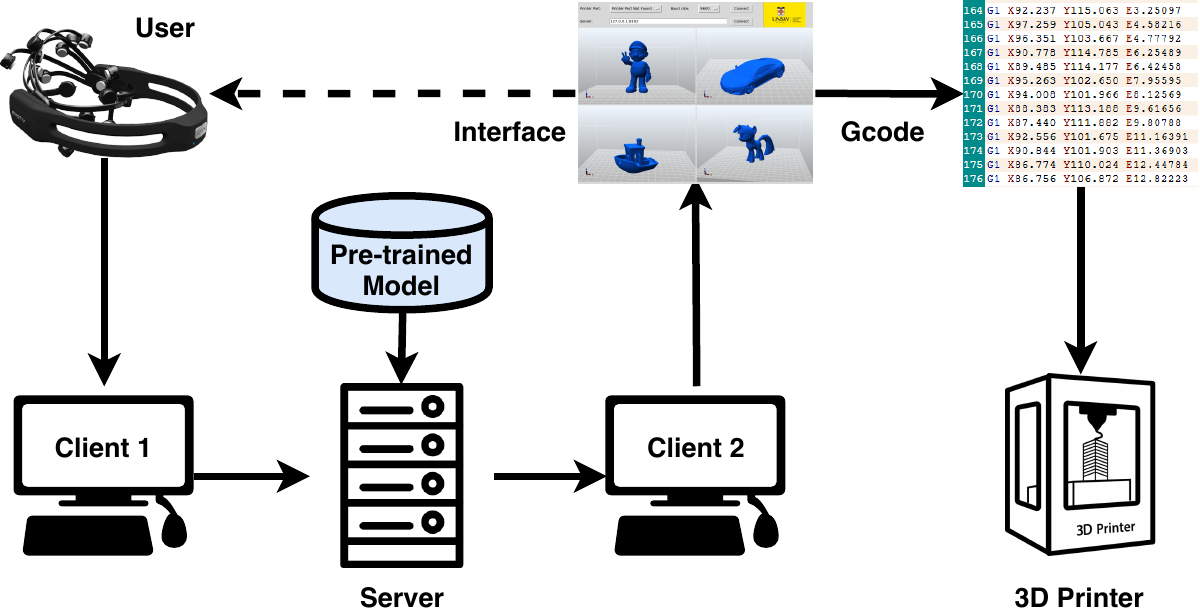}
\caption{Online workflow of {\it Brain2Object}. The user's EEG signals are collected and send to the server through client 1. The server loads the pre-trained model to recognize the target object and send to both the interface for showing the user feedback and the 3D printer for printing. The solid line denotes signal transmission while the slash line denotes feedback.}
\label{fig:online_workflow}
\vspace{-3mm}
\end{figure}

\section{Online Demonstration} 
\label{sec:online_demonstration}
In this section, we summarize our experience in developing a working prototype of Brain2Object. Figure~\ref{fig:online_scenario} shows the working prototype in action. 
The graphical user interface is provided in Figure~\ref{fig:GUI}. The top of the interface shows the port number and baud rate of the IP printer. The IP address of the server which stores the pre-trained model and makes the object recognition decision is also shown.
The main body of the interface displays object models for the four objects in our experiments, namely, Mario, car, boat and Pinkie Pie Pony.

Figure~\ref{fig:online_workflow} illustrates the operational workflow of the Brain2Object demonstrator. 
While the user is focusing on a target object ({\it e.g.}, the Pinkie Pie), the corresponding brain signals are collected by a properly mounted Emotiv headset and transmitted to client 1 over Bluetooth. Client 1 sends the EEG signal to the server over a TCP connection. The server loads the pre-trained EEG recognition model and classifies the EEG signal to one of the four categories. The classification result is forwarded to the interface through client 2. The interface will highlight the selected object by changing the color of the other 3 objects to gray (the selected object remains blue). Simultaneously, the selected object is dispatched to the printer driver which generates the corresponding Gcode which can be recognized by the mechanical 3D printer. Finally, the Gcode is sent to the 3D printer, which brings the object to life. 

We used a Tonxy X1 desktop 3D with the following specifications. Printer size:$220\times 220 \times 250 mm$,  build area: $150\times 150 \times 150 mm$, MK10 extruder diameter: 1.75mm, nozzle diameter: 0.4mm, engraving accuracy: 0.1mm, filament material: 1.75mm polylactic acid (PLA).
The physical 3D model can be transmitted from a computer to the printer or directly stored in minor SD card mounted on the printer. 

The sampling frequency of the Emotiv headset is 128 Hz which indicates it can collect 128 sampling points each second. The pre-trained recognition model requires each sample with 14 sampling point and each sampling point corresponds to a classification output. 
To achieve steadiness and reliability, the server will maintain a window of the last 10 classification output and a count of how many times each of the 4 objects has been recognized. The server will send the target to the client 2 only if one specific target appears more than 6 times in this window. 
\red{There are two subjects participate in the online demonstration. Each subject is tested over 12 trails (3 trails for each object) and we report the average performance. The online system achieves a recognition accuracy of 83.3\% although the latency is increased to about $2$ sec which includes data collection time ($1.1$ sec), recognition time ($0.47$ sec), transmission time, etc.
}



\section{Discussion and Future Work} 
\label{sec:discussion_and_future_work}

In this section, we discuss the challenges and potential directions for future research. 

First, the proposed approach is significantly influenced by the quality of the EEG data.
The pre-trained model shows better performance on the clean and precise public dataset than on our local dataset. This suggests the necessity to develop novel classification methods that are robust to noisy and low-resolution EEG signals.
Another concern is the adaptability over different EEG acquisition equipment. Ideally, the model can consistently achieve accurate performance across a range of hardware platforms. 
 However, the popular platforms ({\it e.g.}, Emotiv\footnote{\red{\url{https://www.emotiv.com/epoc/}}}, NeuroSky\footnote{\red{\url{http://neurosky.com/}}}, OpenBCI\footnote{\red{\url{https://openbci.com/}}}) have different characteristics like sampling resolution, number of channels, channels' position, etc. Thus, there is still innovation required to develop robust and adaptive brain signal classification algorithms. 

Second, the object repository in this work is limited. An ideal instantiation of Brain2Object should recognize any object the user observed. The object repository in this paper only contains four items.
The repository scale is constrained by the learning algorithm, {\it i.e.}, the ability of multi-class classification algorithms to discriminate between a large number of classes. The classification accuracy reduced dramatically with the increase of category numbers. In our pre-experiment which are not presented in this paper due to space limitation, in the offline test, the proposed approach can achieve around 90\% on binary classification using the Emotiv headset, however, the accuracy drops to nears 80\% with three categories and about 70\% with 4 objects. In our future work, we attempt to propose an algorithm to increase the multi-class classification performance. 

Additionally, the ideal printing system is supposed to automatically detects the object which the subject `thinking' (without visual stimulation) instead of `observing' (with visual stimulation). however, the EEG SNR without visual stimulation is much lower than the SNR with visual stimulation. To enhance the SNR and help the subject to concentrate on the object, we adopt visual stimuli in our experiments. Therefore, in the local dataset and the online demonstration phase, the corresponding object images are shown on the monitor to remind the participants. However, the public dataset not only contains visual stimuli but also includes motor imagery, which is one possible reason why EEG-P is classified so accurately. Most of the existing public EEG dataset with visual stimulation focus on motor imagery (like the selected EEG-P) or evoked potentials \cite{odom2010iscev}. In the latter instance, the visual images are flashed at a high frequency which results in pulsed EEG data, {\it i.e.} as objects flash by, short pulses of corresponding EEG data are generated.
 Nevertheless, the model used in this paper is based on table EEG signals caused by steady stimuli. Hence, we select EEG-P instead of evoked potential-based dataset for the evaluation. 

Furthermore, through the online demonstration experiment, we observed that the online performance is lower than the off-line analysis, which could be attributed to several reasons: 1) the user’s mental state and fluctuations in emotions may affect the quality of the EEG signals. For instance, if the pre-trained model is tuned based on the EEG data which is collected when the user is relaxed, the classification performance may be affected while the user is excited in the online phase. 2) the conductive of the electrodes in the headset is not exactly invariant during the off-line stage and online stage, which will have an impact on the data quality; 3) subtle variations ({\it e.g.}, the position of each of the electrodes) in the way the EEG headset is mounted on the subject’s head may also influent online decision making; 4) subjects often have difficulty in maintaining concentration during signal acquisition.

\section{Related Work} 
\label{sec:related_work}
The problem of accurately classifying EEG signals has been extensively researched in recent years  \cite{michelmann2018data,li2015autoregressive,zhang2017classification,liu2017multi,li2015feature,lin2016classification,acharya2018deep,ma2015resting,lawhern2016eegnet,li2017emotion}. Michelmann {\it et al.} \cite{michelmann2018data} use Independent Component Analysis to derive filter coefficients that reflect the statistical dependencies of the data at hand. This work quantitatively shows that ICA outperforms the bipolar referencing operation in sensitivity and importantly in specificity when revealing local time series from the superposition of neighboring channels. Liu {\it et al.} \cite{liu2017multi} propose a multi-layer RNN based model with two attention mechanisms, which achieves high accuracy and boosts the computational efficiency of EEG signal classification. Zhang {\it et al.} \cite{zhang2017classification} present two combination strategies of feature extraction on EEG signals where the first strategy assembles the autoregressive coefficients and approximate entropy and the second strategy adopts wavelet packet decomposition. 

Among the existing studies, the spatial feature-based algorithms had been illustrated to be one of the most promising methods. In particular, CNN, as one of the most popular and powerful deep learning structures, had been widely used to cope with various EEG-based applications. Acharya {\it et al.} \cite{acharya2018deep} employ a standard CNN to analyze patients' brain signals to detect normal, preictal, and the seizure state. Ma {\it et al.} \cite{ma2015resting} apply CNN to automatically extract an individual's best and most unique neural features and conduct classification, using EEG data derived from both resting states with open eyes and resting state with closed eyes, for the aim of individual identification. Lawhern {\it et al.} \cite{lawhern2016eegnet} introduce the use of depth-wise and separable convolutions to construct an EEG-specific model that encapsulates well-known EEG feature extraction concepts for BCI. Zhang {\it et al.} \cite{zhang2018converting} exploit the temporal-spatial information extracted by a combined RNN and CNN, however, this work is based on single sample classification which causes the unstable performance in the online stage.

\section{Conclusion} 
\label{sec:conclusion}
In this paper, we propose an end-to-end printing system based on the combination of multi-class CSP and graph embedded CNN. The performance of the proposed model is evaluated over two datasets in off-line and also demonstrated in the online environment. {\it Brain2Object} serves as a harbinger for exciting BCI applications which can help individuals with various tasks in their daily lives. 
The proposed system employs multi-class CSP to map the EEG data to a common space for the aim of maximizing the distance among various EEG patterns. The processed data are embedded by dynamic graph transformation and then fed into a designed convolutional neural network for automatically spatial feature learning.
Extensive evaluations using a large-scale public dataset and a more relevant but limited local dataset showed that our scheme significantly outperforms a number of state-of-the-art approaches. The system latency is shown to be acceptable and a visualization of the signals is presented to offer additional perspectives into the performance.
The online demonstration is presented to show the applicability of the proposed system.

\section*{References}

\bibliography{TBCI}

\begin{thebibliography}{10}
\expandafter\ifx\csname url\endcsname\relax
  \def\url#1{\texttt{#1}}\fi
\expandafter\ifx\csname urlprefix\endcsname\relax\def\urlprefix{URL }\fi
\expandafter\ifx\csname href\endcsname\relax
  \def\href#1#2{#2} \def\path#1{#1}\fi

\bibitem{tomida2015active}
N.~Tomida, T.~Tanaka, S.~Ono, M.~Yamagishi, H.~Higashi, Active data selection
  for motor imagery eeg classification, IEEE Transactions on Biomedical
  Engineering 62~(2) (2015) 458--467.

\bibitem{zhang2017intent}
X.~Zhang, L.~Yao, C.~Huang, Q.~Z. Sheng, X.~Wang, Intent recognition in smart
  living through deep recurrent neural networks, in: International Conference
  on Neural Information Processing, Springer, 2017, pp. 748--758.

\bibitem{pinheiro2016wheelchair}
O.~R. Pinheiro, L.~R. Alves, M.~Romero, J.~R. de~Souza, Wheelchair simulator
  game for training people with severe disabilities, in: Technology and
  Innovation in Sports, Health and Wellbeing (TISHW), International Conference
  on, IEEE, 2016.

\bibitem{kaya20141d}
Y.~Kaya, M.~Uyar, R.~Tekin, S.~Y{\i}ld{\i}r{\i}m, 1d-local binary pattern based
  feature extraction for classification of epileptic eeg signals, Applied
  Mathematics and Computation 243 (2014) 209--219.

\bibitem{Alomari2014}
M.~H. Alomari, A.~Abubaker, A.~Turani, A.~M. Baniyounes, A.~Manasreh, {EEG
  Mouse : A Machine Learning-Based Brain Computer Interface} 5~(4) (2014)
  193--198.

\bibitem{hamadicharef2010brain}
B.~Hamadicharef, M.~Xu, S.~Aditya, Brain-computer interface (bci) based musical
  composition, in: 2010 International Conference on Cyberworlds, IEEE, 2010,
  pp. 282--286.

\bibitem{pinegger2017composing}
A.~Pinegger, H.~Hiebel, S.~C. Wriessnegger, G.~R. M{\"u}ller-Putz, Composing
  only by thought: Novel application of the p300 brain-computer interface, PloS
  one 12~(9) (2017).

\bibitem{funk2013brain}
M.~Funk, M.~Raschke, Brain painting: Action paintings based on bci-input,
  Mensch \& Computer 2013-Workshopband (2013).

\bibitem{george2010brain}
H.~George, A.~Hosle, D.~Franz, A.~Kubler, Brain painting--bci meets patients
  and artists in the field, in: Integrating Brain-Computer Interfaces with
  Conventional Assistive Technology, TOBI Workshop, 2010, p.~43.

\bibitem{volker2018deep}
M.~V{\"o}lker, R.~T. Schirrmeister, L.~D. Fiederer, W.~Burgard, T.~Ball, Deep
  transfer learning for error decoding from non-invasive eeg, in:
  Brain-Computer Interface (BCI), 2018 6th International Conference on, IEEE,
  2018, pp. 1--6.

\bibitem{jayakar2016diagnostic}
P.~Jayakar, J.~Gotman, A.~S. Harvey, A.~Palmini, L.~Tassi, D.~Schomer,
  F.~Dubeau, F.~Bartolomei, A.~Yu, P.~Kr{\v{s}}ek, et~al., Diagnostic utility
  of invasive eeg for epilepsy surgery: indications, modalities, and
  techniques, Epilepsia 57~(11) (2016) 1735--1747.

\bibitem{zhang2018converting}
X.~Zhang, L.~Yao, Q.~Z. Sheng, S.~S. Kanhere, T.~Gu, D.~Zhang, Converting your
  thoughts to texts: Enabling brain typing via deep feature learning of eeg
  signals, in: 2018 IEEE International Conference on Pervasive Computing and
  Communications (PerCom), IEEE, 2018, pp. 1--10.

\bibitem{shaw2016statistical}
L.~Shaw, A.~Routray, Statistical features extraction for multivariate pattern
  analysis in meditation eeg using pca, in: Student Conference (ISC), 2016 IEEE
  EMBS International, IEEE, 2016, pp. 1--4.

\bibitem{michelmann2018data}
S.~Michelmann, M.~S. Treder, B.~Griffiths, C.~Kerr{\'e}n, F.~Roux, M.~Wimber,
  D.~Rollings, V.~Sawlani, R.~Chelvarajah, S.~Gollwitzer, et~al., Data-driven
  re-referencing of intracranial eeg based on independent component analysis
  (ica), Journal of neuroscience methods 307 (2018) 125--137.

\bibitem{li2015autoregressive}
P.~Li, X.~Wang, F.~Li, R.~Zhang, T.~Ma, Y.~Peng, X.~Lei, Y.~Tian, D.~Guo,
  T.~Liu, et~al., Autoregressive model in the lp norm space for eeg analysis,
  Journal of neuroscience methods 240 (2015) 170--178.

\bibitem{zhang2017classification}
Y.~Zhang, B.~Liu, X.~Ji, D.~Huang, Classification of eeg signals based on
  autoregressive model and wavelet packet decomposition, Neural Processing
  Letters 45~(2) (2017) 365--378.

\bibitem{li2017emotion}
Z.~Li, X.~Tian, L.~Shu, X.~Xu, B.~Hu, Emotion recognition from eeg using rasm
  and lstm, in: International Conference on Internet Multimedia Computing and
  Service, Springer, 2017, pp. 310--318.

\bibitem{liu2017multi}
J.~Liu, Y.~Su, Y.~Liu, Multi-modal emotion recognition with temporal-band
  attention based on lstm-rnn, in: Pacific Rim Conference on Multimedia,
  Springer, 2017, pp. 194--204.

\bibitem{li2015feature}
J.~Li, Z.~Struzik, L.~Zhang, A.~Cichocki, Feature learning from incomplete eeg
  with denoising autoencoder, Neurocomputing 165 (2015) 23--31.

\bibitem{lin2016classification}
Q.~Lin, S.-q. Ye, X.-m. Huang, S.-y. Li, M.-z. Zhang, Y.~Xue, W.-S. Chen,
  Classification of epileptic eeg signals with stacked sparse autoencoder based
  on deep learning, in: International Conference on Intelligent Computing,
  Springer, 2016, pp. 802--810.

\bibitem{simpson2011exponential}
S.~L. Simpson, S.~Hayasaka, P.~J. Laurienti, Exponential random graph modeling
  for complex brain networks, PloS one 6~(5) (2011).

\bibitem{agosta2013brain}
F.~Agosta, S.~Sala, P.~Valsasina, A.~Meani, E.~Canu, G.~Magnani, S.~F. Cappa,
  E.~Scola, P.~Quatto, M.~A. Horsfield, et~al., Brain network connectivity
  assessed using graph theory in frontotemporal dementia, Neurology 81~(2)
  (2013) 134--143.

\bibitem{elisha2017eeg}
A.~E. Elisha, L.~Garg, O.~Falzon, G.~Di~Giovanni, Eeg feature extraction using
  common spatial pattern with spectral graph decomposition, in: Computing
  Networking and Informatics (ICCNI), 2017 International Conference on, IEEE,
  2017, pp. 1--8.

\bibitem{chin2009multi}
Z.~Y. Chin, K.~K. Ang, C.~Wang, C.~Guan, H.~Zhang, Multi-class filter bank
  common spatial pattern for four-class motor imagery bci, in: 2009 Annual
  International Conference of the IEEE Engineering in Medicine and Biology
  Society, IEEE, 2009, pp. 571--574.

\bibitem{kang2014bayesian}
H.~Kang, S.~Choi, Bayesian common spatial patterns for multi-subject eeg
  classification, Neural Networks 57 (2014) 39--50.

\bibitem{wang2006common}
Y.~Wang, S.~Gao, X.~Gao, Common spatial pattern method for channel selelction
  in motor imagery based brain-computer interface, in: Engineering in medicine
  and biology society, 2005. IEEE-EMBS 2005. 27th Annual international
  conference of the, IEEE, 2006, pp. 5392--5395.

\bibitem{song2018eeg}
T.~Song, W.~Zheng, P.~Song, Z.~Cui, Eeg emotion recognition using dynamical
  graph convolutional neural networks, IEEE Transactions on Affective Computing
  (2018).

\bibitem{ciresan2011convolutional}
D.~C. Ciresan, U.~Meier, L.~M. Gambardella, J.~Schmidhuber, Convolutional
  neural network committees for handwritten character classification, in:
  Document Analysis and Recognition (ICDAR), 2011 International Conference on,
  IEEE, 2011, pp. 1135--1139.

\bibitem{ning2018deepmag}
R.~Ning, C.~Wang, C.~Xin, J.~Li, H.~Wu, Deepmag: Sniffing mobile apps in
  magnetic field through deep convolutional neural networks, in: 2018 IEEE
  International Conference on Pervasive Computing and Communications (PerCom),
  IEEE, 2018, pp. 1--10.

\bibitem{ren2017faster}
S.~Ren, K.~He, R.~Girshick, J.~Sun, Faster r-cnn: towards real-time object
  detection with region proposal networks, IEEE Transactions on Pattern
  Analysis \& Machine Intelligence~(6) (2017) 1137--1149.

\bibitem{meisheri2018multiclass}
H.~Meisheri, N.~Ramrao, S.~Mitra, Multiclass common spatial pattern for eeg
  based brain computer interface with adaptive learning classifier, arXiv
  preprint arXiv:1802.09046 (2018).

\bibitem{ramoser2000optimal}
H.~Ramoser, J.~Muller-Gerking, G.~Pfurtscheller, Optimal spatial filtering of
  single trial eeg during imagined hand movement, IEEE transactions on
  rehabilitation engineering 8~(4) (2000) 441--446.

\bibitem{ang2012filter}
K.~K. Ang, Z.~Y. Chin, C.~Wang, C.~Guan, H.~Zhang, Filter bank common spatial
  pattern algorithm on bci competition iv datasets 2a and 2b, Frontiers in
  neuroscience 6 (2012) 39.

\bibitem{yang2018automatic}
B.~Yang, K.~Duan, C.~Fan, C.~Hu, J.~Wang, Automatic ocular artifacts removal in
  eeg using deep learning, Biomedical Signal Processing and Control 43 (2018)
  148--158.

\bibitem{sturm2016interpretable}
I.~Sturm, S.~Lapuschkin, W.~Samek, K.-R. M{\"u}ller, Interpretable deep neural
  networks for single-trial eeg classification, Journal of neuroscience methods
  274 (2016) 141--145.

\bibitem{yang2015use}
H.~Yang, S.~Sakhavi, K.~K. Ang, C.~Guan, On the use of convolutional neural
  networks and augmented csp features for multi-class motor imagery of eeg
  signals classification, in: Engineering in Medicine and Biology Society
  (EMBC), 2015 37th Annual International Conference of the IEEE, IEEE, 2015,
  pp. 2620--2623.

\bibitem{park2014augmented}
C.~Park, C.~C. Took, D.~P. Mandic, Augmented complex common spatial patterns
  for classification of noncircular eeg from motor imagery tasks, IEEE
  Transactions on neural systems and rehabilitation engineering 22~(1) (2014)
  1--10.

\bibitem{thomas2017eeg}
K.~P. Thomas, N.~Robinson, A.~P. Vinod, Eeg-based motor imagery classification
  using subject-specific spatio-spectral features, in: 2017 IEEE International
  Conference on Systems, Man, and Cybernetics (SMC), IEEE, 2017, pp.
  2302--2307.

\bibitem{odom2010iscev}
J.~V. Odom, M.~Bach, M.~Brigell, G.~E. Holder, D.~L. McCulloch, A.~P. Tormene,
  et~al., Iscev standard for clinical visual evoked potentials (2009 update),
  Documenta ophthalmologica 120~(1) (2010) 111--119.

\bibitem{acharya2018deep}
U.~R. Acharya, S.~L. Oh, Y.~Hagiwara, J.~H. Tan, H.~Adeli, Deep convolutional
  neural network for the automated detection and diagnosis of seizure using eeg
  signals, Computers in biology and medicine 100 (2018) 270--278.

\bibitem{ma2015resting}
L.~Ma, J.~W. Minett, T.~Blu, W.~S. Wang, Resting state eeg-based biometrics for
  individual identification using convolutional neural networks, in:
  Engineering in Medicine and Biology Society (EMBC), 2015 37th Annual
  International Conference of the IEEE, IEEE, 2015, pp. 2848--2851.

\bibitem{lawhern2016eegnet}
V.~J. Lawhern, A.~J. Solon, N.~R. Waytowich, S.~M. Gordon, C.~P. Hung, B.~J.
  Lance, Eegnet: A compact convolutional network for eeg-based brain-computer
  interfaces, arXiv preprint arXiv:1611.08024 (2016).

\end{thebibliography}

\end{document}